\documentclass[12pt]{iopart}

\usepackage{graphicx}

\begin{document}

\title{The dispersion modification of electrostatic geodesic acoustic mode by electron geodesic drift current }

\author{Shuangxi Zhang}

\address{Department of Modern Physics, University of Science and Technology of China, Hefei, 230026, China
}
\ead{zshuangxi@gmail.com}

\date{\today}

\begin{abstract}
The past studies treated the perturbed distribution of circulating electrons as adiabatic one when studying the dispersion relation of electrostatic geodesic acoustic mode(GAM). In this paper, the flow of electron geodesic current (FEGC) is added to modify this adiabatic distribution. Based on the drift kinetic theory, it is found that FEGC obviously increases the magnitude of the standard GAM's frequency and reduces its damping rate. The increase of frequency results from the contribution of FEGC to the radial  flow. The reason for the reduction of damping rate is that when the effect of FEGC counts, the new resonant velocity becomes much larger than ions thermal velocity  with equilibrium distribution obeying Maxwellian distribution, compared with unmodified Landau resonant velocity. Especially, FEGC changes the characters of the frequency and damping rate of low-frequency GAM as functions of safety factor $q$ .
\end{abstract}

\pacs{2.25.Dg, 52.30.Gz, 52.25.Xz, 52.55.Fa}
\maketitle

\section{INTRODUCTION}\label{sec1}

Geodesic acoustic mode (GAM) is a well-known experimental phenomena observed in tokamak plasmas\cite{2005conway1,2005hamada1,2006ido1,2008lantao1}. Experiments found that the perturbed electric potential of GAM is symmetrical in toloidal and poloidal direction, whiles the density as the response to the electrostatic perturbation has toroidal number $n=0$ and poloidal $m=1$\cite{2005diamond1,2010qiuzhiyong1,2009fujisawa1}. Recently GAM and zonal flow attract increasing attention due to its ability in reducing turbulent anomalous transport by decreasing the scale of turbulent vortex through shear decorrelation\cite{2005diamond1,1990biglari1,1999itoh1}. Compared with zonal flow, the frequency of GAM is much higher, which ensure its weaker effect on turbulence \cite{2012zhangyangzhong1}. However, experiment studies suggest that GAM is related to L-H transition and transport barriers\cite{2010miki1,2011conway1,2012xumin1}.

GAM, firstly discovered by Winsor et al\cite{1968winsor1} based on magnetic fluid theory, was found to be originated from the compressibility of $\mathbf{E}\times\mathbf{B}$ drift flow, which is proved by the experiment results\cite{2009fujisawa1} that GAM has a $m=1$ accumulated perturbed density in poloidal direction. The past studies divided GAM into electrostatic and electromagnetic kinds. In the studies of electrostatic GAM, past studies, such as Re.(\cite{2005watari1}), equaled circulating electrons' perturbed distribution to the adiabatic distribution\cite{2006sugama1,2006gaozhe1} since it is considered that the transit frequency of circulating electrons is much larger than the oscillating frequency of GAM. However, in the studies of electromagnetic GAM, the effect enforced by nonadiabatic  distribution of circulating electrons on electromagnetic GAM's dispersion is derived\cite{2008smolyakov1,2007zhouden1}. In this paper, the effect of FEGC, which contributes adiabatic response of circulating electrons to GAM's dispersion,  will be studied. The trapped electrons effect and model coupling effect will be ignored in our study. For detail information of trapped electrons, Ref.(\cite{2013zhangshuangxi1,2010zhanghuashen1}) are recommended. We first derive the quasi-neutral equation.

The fact that experiments haven't found perturbed density accumulation in radial direction indicates the balance between all radial flows\cite{2005conway1,2005hamada1,2006sugama1} when GAM exists. Flows in radial direction include geodesic curvature drift flow and polarization drift flow of electrons and ions. The quasi-neutral equation or balance equation of radial flows can be written as
\begin{equation}\label{e1}
\int {Rd\theta {d^3}v\left( {{v_{di}}{{\tilde f}_i}\sin \theta  - {v_{pi}}{F_{0i}} -{v_{de}}{{\tilde f}_e}\sin \theta } \right)}  = 0
\end{equation}
where $\tilde{f}_i$,$\tilde{f}_e$ are the total perturbed distributions for ions and electrons,  $v_{pi(e)}=\omega k\tilde{\phi}/\Omega_{i(e)}B$ is the polarization drift velocity, $v_{de},v_{di}$ are respective radial drift velocity of electrons and ions, which will be defined below. In Eq.(\ref{e1}), the polarization drift flow of electrons is neglected. We will compare the contribution of  term $v_{di}\tilde{f}_i$ to Eq.(\ref{e1}) with  that of $v_{de}\tilde{f}_e$. Compared with $\tilde{f}_e$, $\tilde{f}_i$ suffers an extra effect resulting from the Landau resonance between the frequency of GAM and the transit frequency of circulating ions. We first make a comparison between $\tilde{f}_e$ and the rest part of $\tilde{f}_{i}$ with the Landau resonance part excluded. As a beginning, drift kinetic theory is introduced.

Drift kinetic theory is a simplified edition of gyrokinetic theory by neglecting finite radius effect under the condition of $k\rho\ll 1$ where $k$ is the wave number of the perturbed mode and $\rho$ is the Larmor radius. As for GAM, drift kinetic theory is plausible because of $k_{GAM}\rho\ll 1$. We adopt the drift kinetic equation used in Ref.(\cite{2006sugama1,2006gaozhe1}). To begin, the $k$-th fourier component of the response of perturbed distribution  of circulating particles for the electrostatic perturbed potential of GAM can be divided to be ${\tilde f_{pk}} =  - p{F_{p0}}{\tilde \phi _k}/{T_p} + {\tilde h_{pk}}$ where $p$ is the kind of electric charge. The drift kinetic equation satisfied by $\tilde{h}_{pk}$ is
\begin{equation}\label{e2}
\left( {\omega  - {\omega _{dp}}\sin \theta  + i{\omega _{tp}}\frac{\partial }{{\partial \theta }}} \right){{\tilde h}_{pk}} = \frac{{{eF_{0p}}}}{{{T_p}}}\omega \tilde \phi_{k}
\end{equation}
where geodesic acoustic frequency ${\omega _{dp}} = k_\perp(v_ \bot ^2/2 + v_\parallel ^2)/{{2{\Omega_p}R}}$, transit frequency ${\omega _{tp}} = {v_\parallel }/qR $, and $k_{\perp}$ is the wave vector of GAM perpendicular to equilibrium magnetic field. $q$ is the safety factor. The poloidal part of  $k_{\perp}$ for the perturbed function of GAM is too small to be included so that  $k_{\perp}=k_r$. The dynamics of circulating electrons and ions obeys Eq.(\ref{e2}). What makes the perturbed distribution of ions differ from adiabatic one is the $\omega_{ti}$ and $\omega_{di}$ terms. $\omega_{ti}$ is relevant to the part of $\tilde{f}_i$ resulting from Landau resonance. $\omega_{di}$ is related the other part of $\tilde{f}_i$  which we want to compare with $\tilde{f}_e$. For circulating electrons, $\omega_{te}$ can be neglected since it's so larger than $\omega$  that resonance can't happen. The only term makes the perturbed distribution of electrons differ from adiabatic one is the $\omega_{te}$ term. The past studies simply neglected this term, but it's found that $\omega_{de}$ is of the same order as $\omega_{di}$ that  the effect $\omega_{de}$  enforces on $\tilde{f}_e$ can be compared to that $\omega_{de}$ does on $\tilde{f}_i$.

 To compare  the  quantity $v_{de}$ with $v_{di}$, we will begin with their definitions. In this paper, magnetic field of large aspect ratio and circular cross section, that's ${\bf{B}} = {B_0}\left( {{{\bf{e}}_\xi } + {{\bf{e}}_\theta }\varepsilon/q} \right)/\left( {1 + \varepsilon\cos \theta } \right)$, is adopted, where $\varepsilon=r/R$ is assumed to be a small number. $\xi,\theta$ are toroidal and poloidal angles, respectively. The magnetic field's curvature is $\kappa  = \left( {\sin \theta {{\bf{e}}_\theta } - \cos \theta {{\bf{e}}_r}} \right)/R$. The $\sin\theta$ part of curvature is called geodesic curvature which is the origin to make $\mathbf{E}\times\mathbf{B}$ drift flow compressible. The geodesic drift velocity is ${{\mathbf{v}}_{dp}} =  - m\left( {v_\parallel ^2 + v_ \bot ^2/2} \right)\sin \theta {{\mathbf{e}}_r}/p{B_0}{R_0}$ in the magnetic field profile given before, where $m$  and $Q$ are particles' mass and electric charge. The integrals of the geodesic drift current of electrons and ions in Maxwellian distribution of velocity  space  becomes
  \begin{equation}\label{e3}
 - \left\langle {{v_{di}}} \right\rangle \sin \theta  =  - \left\langle {\frac{{\left( {v_ \bot ^2/2 + v_\parallel ^2} \right)}}{{{\omega _{ci}}R}}} \right\rangle \sin \theta  =  - \frac{{{T_i}}}{{RB}}\sin \theta,
  \end{equation}
\begin{equation}\label{e4}
 - \left\langle {{v_{de}}} \right\rangle \sin \theta  = \left\langle {\frac{{\left( {v_ \bot ^2/2 + v_\parallel ^2} \right)}}{{{\omega _{ce}}R}} } \right\rangle \sin \theta {\rm{ = }}\frac{{{T_e}}}{{RB}}\sin \theta.
\end{equation}
The geodesic drift currents of electrons and ions have the same average magnitude but opposite direction under the assumption of thermal equilibrium between ions and electrons.

Based on the discussion before, we found that the part of $v_{de}\tilde{f}_e$ from the effect of $\omega_{de}$ term has the same order as the part of $v_{di}\tilde{f}_i$ from the effect of $\omega_{di}$ term, which introduces the modification to the adiabatic distribution of passing electrons. In the rest of the paper, we will get a analytical study of the effect enforced by FEGC on GAM's dispersion and make a comparison with that of no such term effect.

In this paper, the effects of finite  radius and model coupling on GAM's dispersion are neglected by focusing on electrons' geodesic drift flow effect. And it's observed from Eq.(\ref{e2}) that when the compressibility  of $\mathbf{E}\times\mathbf{B}$ drift flow is fully compensated by an ion acoustic parallel return flow, a nature solution of $\omega$ is $0$.  Such a solution is just a simple example of Liouville's theorem and corresponds to zonal flow. If the two terms can't  cancel out each other, then $\omega  \ne 0$, which leads to the appearance of GAM. As experiments\cite{2009fujisawa1} and theories\cite{2006gaozhe1} reveal, GAM may have more than one branch. The reference\cite{2006gaozhe1} found GAM has two branches, the standard branch and the low frequency branch. We will study the two branches' dispersion relation, respectively.

The arrangement of the rest paper is as follows. In Sec.(\ref{sec2}), the GAM's dispersion function is derived. In Sec.(\ref{sec3}), we make the numerical study of GAM's dispersion relation. Sec.(\ref{sec4}) is the summery and discussion

\section{The derivation of GAM's dispersion equation}\label{sec2}

GAM's drift kinetic equation with magnetic field of large aspect ratio and circular cross section is giving in Eq.(\ref{e2}). $\tilde{\phi}_k, \tilde{h}_k$ can be  decomposed as Fourier series ${\tilde \phi _k} = \sum\limits_{n =  - \infty }^{ + \infty } {{{\tilde \phi }_{k,n}}{e^{in\theta }}} ,{\tilde h_k} = \sum\limits_{n =  - \infty }^{ + \infty } {{{\tilde h}_{k,n}}{e^{in\theta }}} $ in Eq.(\ref{e2}). The authors gave the analytical solution of Eq.(\ref{e2}) in \cite{2006gaozhe1}
\begin{equation}\label{e5}
{{\tilde h}_{p,k}} =   \frac{{p{F_{0p}}\tilde \phi }}{{{T_p}}}\sum\limits_{m,n =  - \infty }^\infty  {{i^{m - n}}{e^{i\left( {m - n} \right)\theta }}\frac{{\omega {J_m}{J_n}}}{{\omega  + \left( {n - l} \right){\omega _{tp}}}}}
\end{equation}
where ${J_{m,n}}$ is a function of $\omega_{dp}/\omega_{tp}$. The right hand of Eq.(\ref{e5}) is a infinite series. $\omega_{dp}$ and $\omega_{tp}$ satisfy the relation $\omega_{dp}/\omega_{tp}\ll 1$ for GAM. Based on the order relationship ${\tilde\phi _{k,m}}/{\tilde\phi _{k,0}} \sim {\left( \omega_{di}/\omega_{ti}\right)^m}$ in sugama's paper\cite{2006sugama1}, $\tilde{\phi}_{k,m}$ can be neglected compared with $\tilde{\phi}_{k,0}$. In this paper, only $\tilde{\phi}_{k,0}$ is kept and modes coupling effect is neglected safely. It's observed from Eq.(\ref{e5}) that only $\sin\theta$ part of $\tilde{f}_i$ and $\tilde{f}_e$ contributes to ions and electrons geodesic drift currents, respectively. This part is obtained from Eq.(\ref{e5})
\begin{equation}\label{e6}
{\tilde h_{p,k}} = \frac{{ p{F_{0p}}{{\tilde \phi }_0}}}{{{T_p}}}\sum\limits_{n = 0}^{{\rm{ + }}\infty } {\left( {\frac{{i\omega {J_{ - n + 1}}{J_{ - n}}}}{{\omega  - n{\omega _t}}}{e^{i\theta }} - \frac{{i\omega {J_{n - 1}}{J_n}}}{{\omega  + n{\omega _t}}}{e^{ - i\theta }}} \right)}.
\end{equation}
Here, terms like ${J_{n + 1}}{J_n},{J_{ - n - 1}}{J_{ - n}},n > 0$ are ignored due to their higher order of $O\left( {\omega _{dp}^2/\omega _{tp}^2} \right)$. Multiplying $\tilde{h}_{p,k}$ by $v_{dp}$ and reformulating the results, we get
\begin{equation}\label{e7}
{v_{dp}}\tilde h = \sum\limits_{n = 0}^{ + \infty } {\frac{{2 p\sqrt n \omega }}{{qk}}J_n^2\left( {\frac{{{v_\parallel }}}{{\omega  - n{\omega _{tp}}}} - \frac{{{v_\parallel }}}{{\omega  + n{\omega _{tp}}}}} \right)\sin \theta }.
\end{equation}
Here, the following properties of Bessel function are used for $|z|<1$
\begin{equation}\label{e8}
{J_{ - n}}\left( z \right) = {\left( { - 1} \right)^n}{J_n}\left( z \right),{J_n}\left( z \right) = {\left( {z/2} \right)^n}/n, \nonumber \\
z{J_n}\left( z \right) = 2\left( {n + 1} \right){J_n}\left( z \right)
\end{equation}
In this paper, only resonance between $\omega$ and $\pm \omega_{tp}$ is considered.

Next we derive the modification resulting from FEGC of circulating electrons' distribution to the adiabatic distribution . With thermal equilibrium between ions and electrons, the thermal velocity of circulating electrons is much larger than that of circulating ions. Landau resonance between GAM and circulating electrons can't happen since $\omega_{GAM}\ll v_e/qR$. As studied in \cite{2013zhangshuangxi1,2010zhanghuashen1}, the effect of trapped electrons on the damping of GAM should not be neglected in high $q$ regime. In this paper, to emphasize the effect of FEGC, the effects of
trapped electron, Landau resonance between circulating electrons transit frequency and GAM's frequency, model coupling, and terms of order higher than $O(\omega_{dp}^{2}/\omega_{tp}^{2})$ are neglected in electrons' drift kientic equation.
 After doing that, the $\sin\theta$ part of non-adiabatic distribution is
\begin{equation}\label{e9}
{{\tilde f}_1} = \frac{{ - p{F_{Mp}}{\bar{\phi _0}}}}{{{T_p}}}\frac{{{\omega _{dp}}}}{\omega }\sin \theta
\end{equation}
The average of $ {\frac{{{v_{dp}}{\omega _{dp}}}}{\omega }} $ in velocity space is
$\left\langle {\frac{{{v_{dp}}{\omega _{dp}}}}{\omega }} \right\rangle  = \frac{7}{4}\frac{{kv_p^4}}{{\omega {R^2}\omega _p^2}}$
where $\left\langle \right\rangle$ means velocity integral in Maxwellian distribution. Substituting of  Eqs.(\ref{e7},\ref{e9}) and polarization drift into Eq.(\ref{e5}), the governing equation is
\begin{equation}\label{e10}
\int_L^{} {\frac{{{d^3}v\exp \left( { - {v^2}} \right)}}{{{\pi ^{3/2}}}}\sum\limits_{n = 0}^{ + \infty } {J_n^2\left( {kq\frac{{2v_\parallel ^2 + v_ \bot ^2}}{{2{v_\parallel }}}} \right)} } \nonumber \\
\times \left( {\frac{{{v_\parallel }}}{{{\zeta _{{i}}}/n - {v_\parallel }}} - \frac{{{v_\parallel }}}{{{\zeta _{{i}}}/n + {v_\parallel }}}} \right) - \frac{{{k^2}}}{2}{\rm{ + }}\frac{{7{k^2}{q^2}}}{{4\zeta _{{e}}^2}} = 0,
\end{equation}
where the following normalization and definition are adopted: $v = {v \mathord{\left/
 {\vphantom {v {{v_{tp}},k = k{v_{tp}}/{\Omega _p},{\zeta _p} \equiv qR\omega /{v_{tp}}}}} \right.
 \kern-\nulldelimiterspace} {{v_{tp}},k = k{v_{tp}}/{\Omega _p},{\zeta _p} \equiv qR\omega /{v_{tp}}}}$ with $v_{tp}=\sqrt{2T_p/m_p}$. compared
  with the GAM's governing equation in Ref.(\cite{2006gaozhe1}), Eq.(\ref{e10}) possesses a term corresponding to FEGC. The sum of $n=0$ terms equals zero in Eq.(\ref{e10})and we keep terms up to order $\omega_{de}^2/\omega_{t}^2$ in the integral for $n=1$ terms. Using the
  following integrating formulas
 \begin{equation}\label{e11}
Z\left( \zeta  \right) = \frac{1}{{\sqrt \pi  }}\int {\frac{{\exp \left( { - v_\parallel ^2} \right)}}{{{v_\parallel } - \zeta }}} d{v_\parallel },
\end{equation}
\begin{equation}\label{e12}
{Z_1}\left( \zeta  \right) = \frac{1}{{\sqrt \pi  }}\int {\frac{{\exp \left( { - v_\parallel ^2} \right)v_\parallel ^2}}{{v_\parallel ^2 - {\zeta ^2}}}} d{v_\parallel } = 1 + \zeta Z\left( \zeta  \right),
\end{equation}
\begin{equation}\label{e13}
{Z_2}\left( \zeta  \right)=\frac{1}{{{\pi ^{3/2}}}}\int {2\pi {v_ \bot }d{v_ \bot }d{v_\parallel }\frac{{\exp \left( { - v_\parallel ^2 - v_ \bot ^2} \right)\left( {v_\parallel ^2 + v_ \bot ^2/2} \right)}}{{v_\parallel ^2 - {\zeta ^2}}}}  \nonumber \\
= \frac{{Z\left( \zeta  \right)}}{{2\zeta }} + {Z_1}\left( \zeta  \right),
\end{equation}
\begin{equation}\label{e14}
{Z_3}\left( \zeta  \right) \nonumber \\
 = \frac{1}{{{\pi ^{3/2}}}}\int {2\pi {v_ \bot }d{v_ \bot }d{v_\parallel }\frac{{\exp \left( { - v_\parallel ^2 - v_ \bot ^2} \right)\left( {v_\parallel ^2 + v_ \bot ^2/2} \right)v_\parallel ^2}}{{v_\parallel ^2 - {\zeta ^2}}}} \nonumber \\
 = 1 + {\zeta ^2}{Z_2}\left( \zeta  \right),
\end{equation}
\begin{equation}\label{e15}
{Z_4}\left( \zeta  \right)  \nonumber \\
 = \frac{1}{{{\pi ^{3/2}}}}\int {2\pi {v_ \bot }d{v_ \bot }d{v_\parallel }\frac{{\exp \left( { - v_\parallel ^2 - v_ \bot ^2} \right){{\left( {v_\parallel ^2 + v_ \bot ^2/2} \right)}^2}}}{{v_\parallel ^2 - {\zeta ^2}}}} \nonumber \\
= \frac{{Z\left( \zeta  \right)}}{{2\zeta }} + \frac{{{Z_1}\left( \zeta  \right)}}{2} + {Z_3}\left( \zeta  \right),
\end{equation}
GAM's governing equation Eq.(\ref{e10}) is simplified to be
\begin{equation}\label{ee3}
\frac{1}{{{q^2}}} - \frac{7}{{8\zeta _i^2}} + \frac{{Z\left( {{\zeta _i}} \right)}}{{2{\zeta _i}}} + \frac{{{Z_1}\left( {{\zeta _i}} \right)}}{2} + {Z_3}\left( {{\zeta _i}} \right) = 0,
\end{equation}
where the ratio $v_{ti}/v_{te}=1/40$ is adopted with thermal equilibrium between ions and electrons. A new term $-7/8\zeta_i^2$ appears in Eq.(\ref{ee3}) compared with the govering relation gotten by the previous papers such as Eq.(5)in Ref.(\cite{2006gaozhe1}).  This term is originated from the including of FEGC term $ {v_{de}}{{\tilde f}_e}\sin \theta$ in the quasi-neutral Eq.(\ref{e1}). The reason is that electron geodesic drift velocity, which has the same amplitude as that for ion geodesic drift velocity, leads to the electron charge density accumulation in the poloidal direction and then introduces electron radial flow
as the diamagnetic flow of the perturbed electron density in poloidal direction to the radial flow balance equation Eq.(\ref{e1}). EGC introduces the modification to the frequency of GAM or to the frequency of reversing the direction of radial flow to keep the quasi-neutral condition. In this model, the damping of GAM comes from the Landau resonance between the frequency of GAM and the transit frequency of ions. So the variation of frequency of GAM should lead to the changing of the damping. These phenomena will be discussed in detail by using the following figures.

\section{The numerical study of GAM's governing equation}\label{sec3}

In this section, we numerically study the modified dispersion equation of GAM Eq.(\ref{ee3}) and make the comparison between it and the old
dispersion relation in Ref.(\cite{2006gaozhe1}). The range of safety factor is chosen to be $(1,5)$. The frequency and damping rate are normalized by $qR/v_{ti}$. It is convenient to transform  the plasma dispersion relation to be the following form for numerical calculating.
\begin{equation}
Z\left( \zeta  \right) = {e^{ - {\zeta ^2}}}\left( {i{\pi ^{1/2}} - 2\int_0^\zeta  {{e^{{x^2}}}dx} } \right).
\end{equation}

\begin{figure}
\includegraphics[height=8cm,width=8cm]{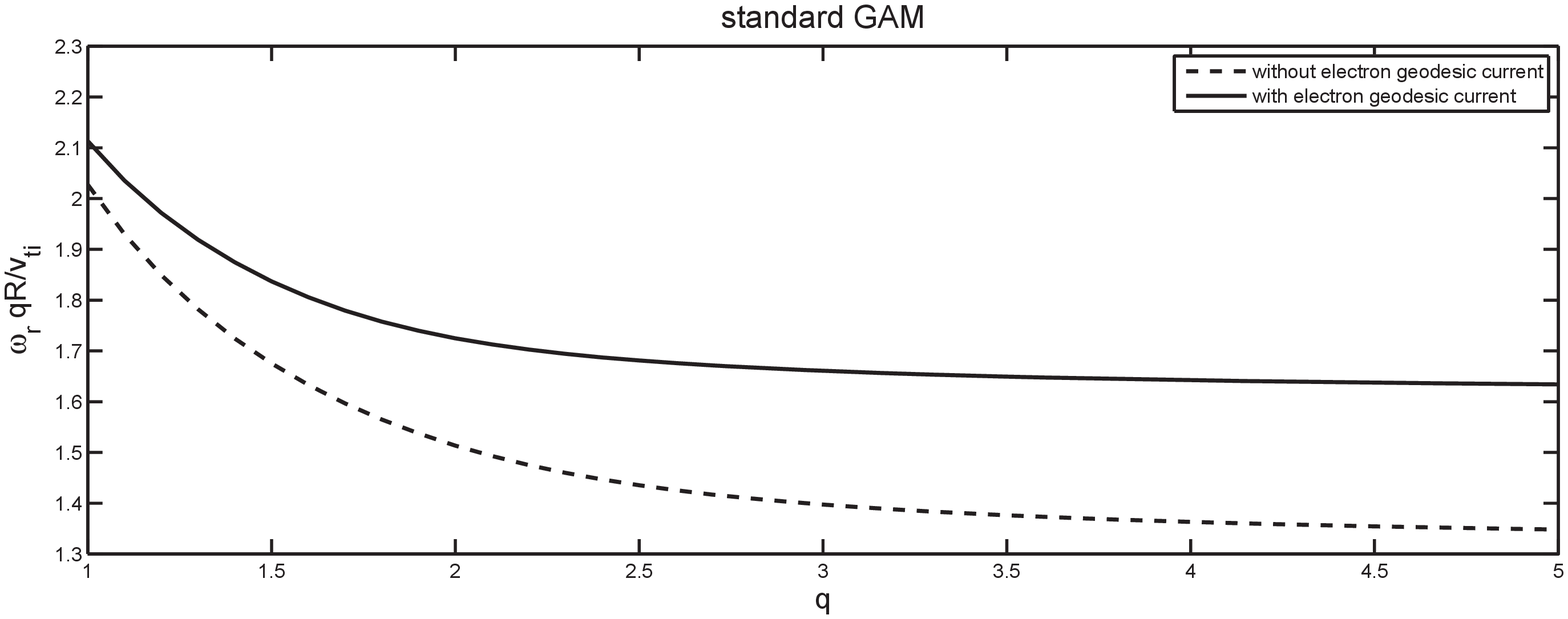}\centering
\caption{\label{comfregamiongam} Comparison of standard GAM's normalized frequency without and with FEGC.}
\end{figure}

\begin{figure}
\includegraphics[height=8cm,width=8cm]{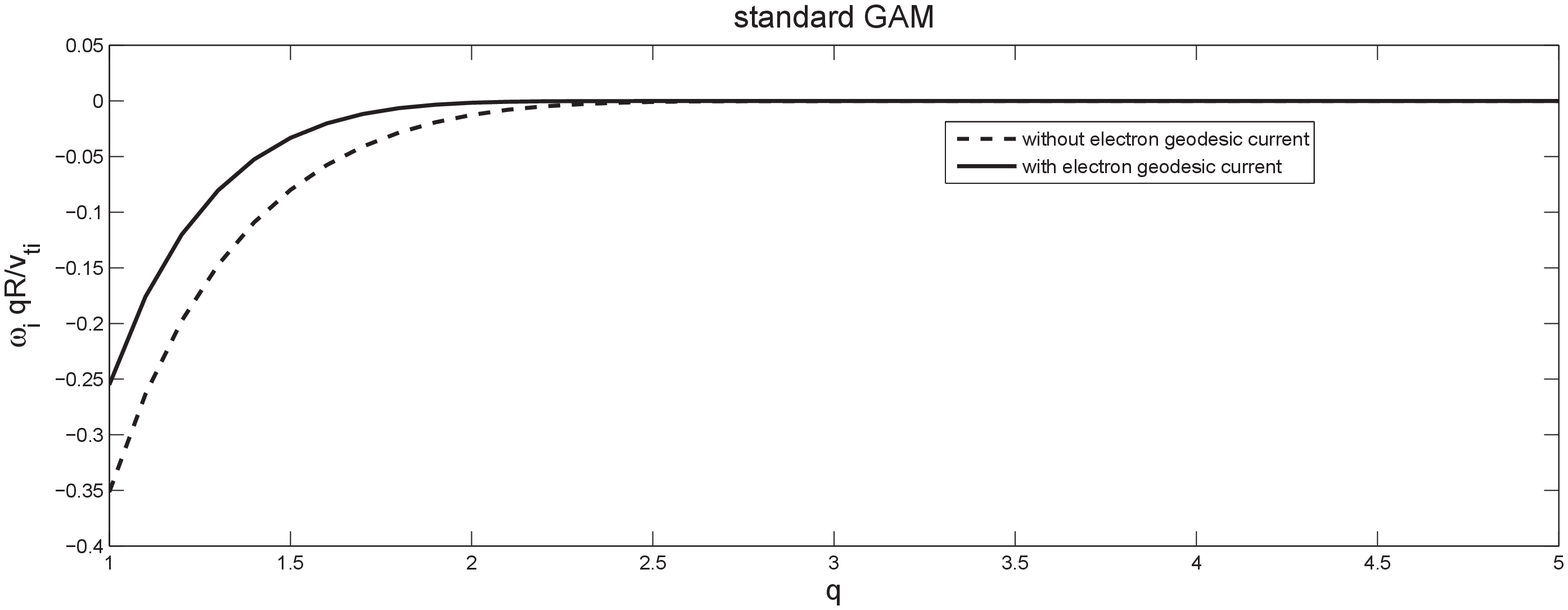}\centering
\caption{\label{comdampgamiongam} Comparison of standard GAM's normalized damping rate without  and with FEGC.}
\end{figure}

The modification of electrons geodesic drift flow on the frequency and damping rate of  standard GAM are shown in Fig.(\ref{comfregamiongam})
and (\ref{comdampgamiongam}). Fig.(\ref{comfregamiongam}) shows that when the FEGC is included, the modified frequency of standard GAM
is higher than the old one \cite{2006gaozhe1}], especially in high $q$ range. Compared with the old one, the magnitude of the new frequency
increases about $1/5$ of that of the old one. The explanation is as follows. As can be seen from Eq.(\ref{e2},\ref{e3}), the geodesic drift flows of ions and electrons are radial flows but have the opposite directions. It is found that approximating to $O(\omega _d/\omega)$, the perturbed densities of electrons and ions have the same sign by comparing the perturbed distribution of electrons in Eq(\ref{e9}) with the perturbed density of ions $\tilde n =  - \frac{{{n_0}{k_r}{{\tilde \phi }_0}\sin \theta }}{{\omega {B_0}{R_0}}}$. Due to the opposite sign between the charge of electron and ion, it can be concluded that the radial charge flow provided by perturbed electron density is in the same direction as that of radial charged flow provided by the perturbed ion density. So the including of FEGC increases the charge flow in the radial flow compared with the old model in Ref(\cite{2006gaozhe1}). In our model, the only flow to balance the radial flow
caused by geodesic curvature is THE polarization flow of ions. Thus to make the balance between radial flow, the frequency of GAM should become bigger, which is more close to the real physics compared with the old model.

As shown in Fig.(\ref{comdampgamiongam}), the modification of the damping rate by FEGC is that the damping rate of GAM is reduced by $1/3$
of its amplitude in the low safety factor range. The reason for the reduction of damping rate can be explained below.
The frequency of GAM increases when FEGC counts, but the only damping rate in our model is the Landau resonance between the
frequency of GAM and the frequency of circulating $\omega_t=v_{\parallel/pR}$. The previous paper
Ref.(\cite{2013zhangshuangxi1})  explains the modification of damping rate by recognizing that the resonant velocity of particle is different from the thermal velocity under Maxwellian  distribution. That paper also pointed out that for three dimensional Maxwellian distribution,
the maximum damping rate happens when the resonant  velocity of particles equals the thermal velocity of particles because the particles has the greatest density when the velocity equals the thermal velocity under  Maxwellian distribution. For circulating ions, due to
resonant condition $\omega_t=\omega$, the increasing of $\omega$  makes the resonant velocity of circulating ions
larger than that of the old model as well as larger than the thermal velocity, thus the number of resonant circulating ions becomes less.
Eventually, the including of electron geodesic flow reduces the damping rate of GAM.

\begin{figure}
\includegraphics[height=8cm,width=8cm]{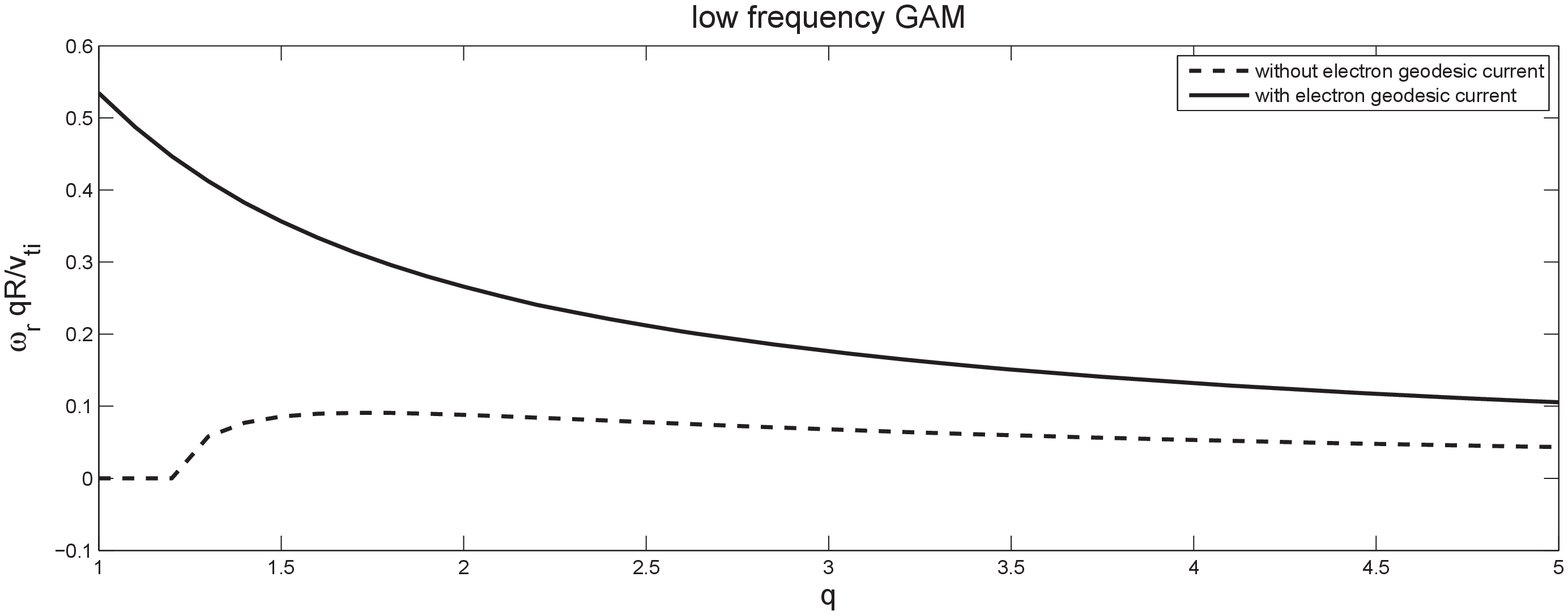}\centering
\caption{\label{comfrezerogamiongam} Comparison of low-frequency GAM's normalized frequency without and with FEGC.}
\end{figure}

\begin{figure}
\includegraphics[height=8cm,width=8cm]{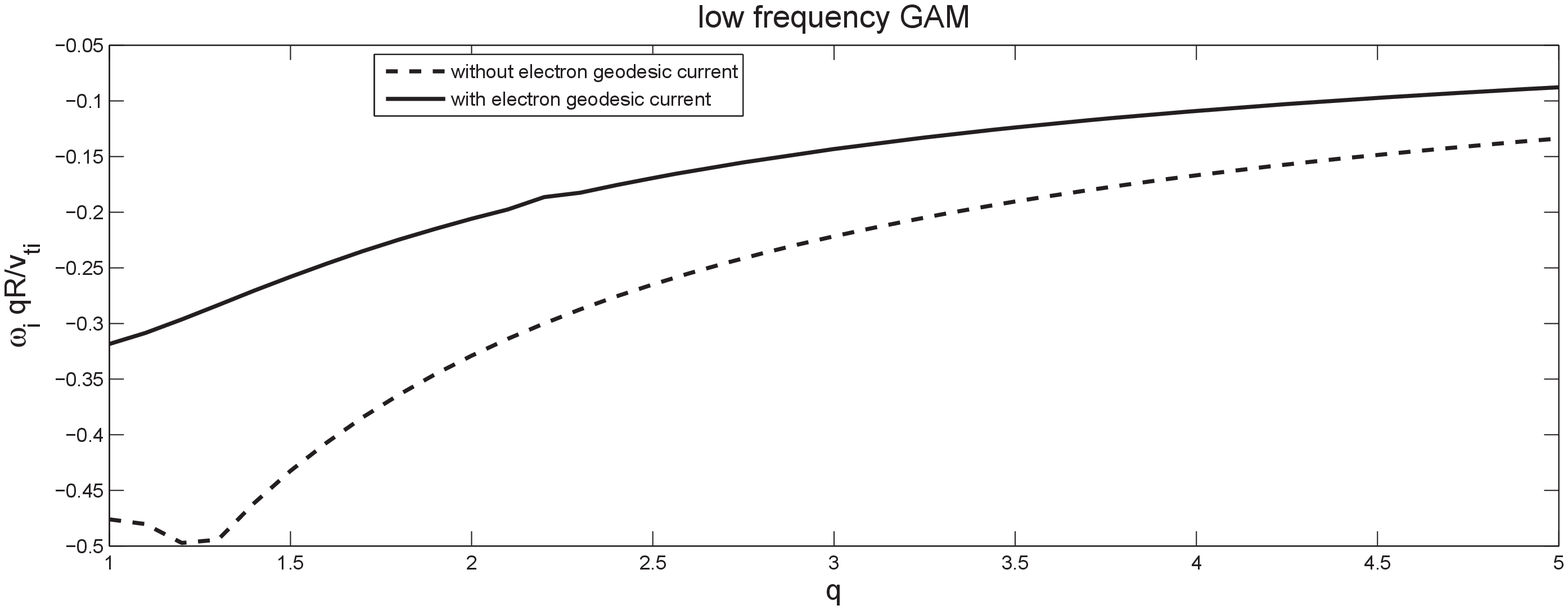}\centering
\caption{\label{comdampzerogamiongam} Comparison of low-frequency GAM's normalized damping rate without and with FEGC.}
\end{figure}

The modification of electrons geodesic drift flow on the frquency and damping rate of low-frequency GAM rate are shown in Fig.(\ref{comfrezerogamiongam}) and (\ref{comdampzerogamiongam}). As Fig.(\ref{comfrezerogamiongam}) shows, with the modification of FEGC, THE frequency of low-frequency GAM is a decreasing function of safety factor and it doesn't equals $0$ when the number of $q$ is close to $1$ as the old frequency does. And in the range of $q\in (1,5)$, the modified frequency is much larger than that without modification especially in  low $q$ range, the reason for which is the same as explained for the standard GAM. So the low-frequency branch of GAM can't appear with such a low frequency as predicted by \cite{2006gaozhe1} in actual physical phenomena. The modification of damping rate of low-frequency GAM is also significant. The amplitude of low-frequency GAM's damping rate reduced by $1/3$ in the low $q$ range as shown in Fig.(\ref{comdampzerogamiongam}). And the peak point of the damping rate without modification disappears after FEGC is included.

It's worthy noting that the dot line in Fig.(\ref{comfrezerogamiongam}) shows that the frequency is too low to be damped by circulating ions.
However, the damping rate shown by the dot line in Fig.(\ref{comfrezerogamiongam}) is even bigger than that of standard GAM. So it may be
inferred that low-frequency GAM is not an eigen model in such kind a geometry and will be studied later.

\section{Summary and Discussion}\label{sec4}

In this paper, we focus on the effect brought by FEGC on the dispersion relation of starand GAM and low-frequency GAM, respectively. It is found that for both branches, FEGC significantly changes the frequency and damping rate compared with those unmodified ones, which has never been discussed before. The reason for the increasing of the frequency of standard GAM is that FEGC increases the radial flow driven by geodesic curvature, thus the balance of radial flow requires a larger radial polarization flow of ions. The reducing of the  damping rate of standard GAM results from the increase of the resonant velocity of circulating ions  due to the increase of oscillating frequency. For a more comprehensive understanding of the mechanism of GAM, the trapped electrons effect and the radial structure of GAM should be included, which is  the future work.

\clearpage
\centering

\clearpage

\section*{References}

\bibliographystyle{pst}
\bibliography{iongamwithelectronflow}

\end{document}